\documentclass[aps,prl,twocolumn,showpacs,superscriptaddress,reprint]{revtex4-1}

\usepackage{graphicx,charter,helvet,sansmath,color}
\usepackage{cancel,dirtytalk,graphicx,blindtext,mathtools}
\usepackage{float}
\usepackage[normalem]{ulem}
\usepackage[utf8]{inputenc}
\graphicspath{ {./Images/} }

\begin{document}

\title{Superconductivity in Compositionally-Complex Cuprates with the YBa$_2$Cu$_3$O$_{7-x}$ Structure}

 \author{Aditya Raghavan}
  \affiliation{Department of Materials Science and Engineering, University of Florida, Gainesville, Florida 32611, USA}
  
 \author{Nathan Arndt}
  \affiliation{Department of Materials Science and Engineering, University of Florida, Gainesville, Florida 32611, USA}
 
 \author{Nayelie Morales-Col\'on}
  \affiliation{Department of Materials Science and Engineering, University of Florida, Gainesville, Florida 32611, USA}
 
  \author{Eli Wennen}
  \affiliation{Department of Materials Science and Engineering, University of Florida, Gainesville, Florida 32611, USA}

  \author{Megan Wolfe}
  \affiliation{Department of Materials Science and Engineering, University of Florida, Gainesville, Florida 32611, USA}

  \author{Carolina Oliveira Gandin}
  \affiliation{Department of Materials Science and Engineering, University of Florida, Gainesville, Florida 32611, USA}

  \author{Kade Nelson}
  \affiliation{Department of Materials Science and Engineering, University of Florida, Gainesville, Florida 32611, USA}

  \author{Robert Nowak}
  \affiliation{Department of Physics, University of Florida, Gainesville, Florida 32611, USA}

  \author{Sam Dillon}
  \affiliation{Department of Physics, University of Florida, Gainesville, Florida 32611, USA}
  
  \author{Keon Sahebkar}
  \affiliation{Department of Materials Science and Engineering, University of Florida, Gainesville, Florida 32611, USA}
  
\author{Ryan F. Need}
 \affiliation{Department of Materials Science and Engineering, University of Florida, Gainesville, Florida 32611, USA}

\begin{abstract}
High-temperature superconductivity is reported in a series of compositionally-complex cuprates with varying degrees of size and spin disorder. Three compositions of Y-site alloyed YBa$_2$Cu$_3$O$_{7-x}$, i.e., (5Y)BCO, were prepared using solid-state methods with different sets of rare earth ions on the Y-site. Synchrotron X-ray diffraction and energy-dispersive X-ray spectroscopy confirm these samples have high phase-purity and homogeneous mixing of the Y-site elements. The superconducting phase transition was probed using electrical resistivity and AC magnetometry measurements, which reveal the transition temperature, T$_C$, is greater than 91 K for all series when near optimal oxygen doping. Importantly, these T$_C$ values are only $\approx$1$\%$ suppressed relative to pure YBCO (T$_C$ = 93 K). This result highlights the robustness of pairing in the YBCO structure to specific types of disorder. In addition, the chemical flexibility of compositionally-complex cuprates allows spin and lattice disorder to be decoupled to a degree not previously possible in high-temperature superconductors. This feature makes compositionally-complex cuprates a uniquely well-suited materials platform for studying proposed pairing interactions in cuprates.
\end{abstract}

\maketitle


Disorder in superconductors has intrigued physicists and been the subject of numerous scientific studies over the last 60 years, starting with Anderson's theory and continuing to present day \cite{anderson_theory_1959, fisher_quantum_1990, pan_microscopic_2001, dubi_nature_2007, sacepe_localization_2011, mizuguchi_glassy_2023}. In s-wave superconductors, Anderson predicted that electron pairing survives non-magnetic disorder in the weak disorder limit \cite{anderson_theory_1959}. Later, Lee and Ma evaluated the strong disorder limit in s-wave systems and predicted increasing disorder would eventually cause a superconducting-insulator transition, as was subsequently observed in amorphous thin films \cite{ma_localized_1985, hebard_magnetic-field-tuned_1990}. By contrast, in unconventional d-wave superconductors, Abrikosov-Gor'kov (AG) theory predicts a universal $T_C$ suppression even for weak non-magnetic disorder, when that disorder is in or near the superconducting plane \cite{abrikosov1960contribution,tolpygo_universal_1996}. However, experimental evidence shows d-wave systems are more robust to disorder located away from the superconducting planes \cite{fujita_effect_2005}.

Recently, a new experimental approach to studying the the interplay of disorder and superconductivity has been enabled by the advent of high-entropy alloys (HEAs) \cite{sun_high-entropy_2019}. HEAs are intermetallic compounds in which several elements (typically four or more) are alloyed together creating materials with large configurational entropy that thermodynamically favors the formation of single phase solid-solutions with complex compositions. In 2014, the first superconducting HEA was reported in the compound Ta$_{34}$Nb$_{33}$Hf$_8$Zr$_{14}$Ti$_{11}$ with a transition temperature, $T_C$, of 7.3 K and Type-II behavior \cite{kozelj_discovery_2014}. Since then, many other superconducting HEAs have been successfully synthesized in a variety of different structure types \cite{von_rohr_effect_2016, sun_high-entropy_2019,hirai_superconductivity_2023}. To date, studies on HEA superconductors collectively demonstrate $T_C$ values between those of amorphous alloys and simple binary alloys, which is consistent with the trend in s-wave systems that increasing disorder tends to suppress pairing and $T_C$ \cite{sun_high-entropy_2019}.

More recently, this entropy-inspired materials engineering approach has been extended to cuprate d-wave systems \cite{musico_synthesis_2021, mazza_searching_2022}. These so-called high-entropy oxides (HEOs) are ionic analogs to HEAs, in which several elements are selectively substituted onto a specific ionic sublattice by judicious choice of ion size and valence \cite{rost_entropy-stabilized_2015}. Thus far, studies of HEO cuprates have only focused on the RE$_2$CuO$_4$ (RCO) system with the Ruddlesden-Popper structure and alloying on the rare earth (RE)-site \cite{musico_synthesis_2021, mazza_searching_2022}, hereafter denoted (5R)CO. Music\'o et al. compared synthetic routes for bulk powders using only trivalent ions and showed both traditional solid-state and sol-gel methods were able to produce well-mixed, single-phase material; however, no attempt was made to charge doped that material into the superconducting phase \cite{musico_synthesis_2021}. The other two studies grew (5R)CO as epitaxial thin films using (5R)CO targets and pulsed laser deposition \cite{zhang_applying_2020, mazza_searching_2022}, and attempts were made to optimally charge dope both (5R)CO films using both hole and electron dopants. However, all samples showed insulating behavior regardless of doping. Using X-ray absorption measurements of the Cu coordination environment, Mazza et al. attributed this result to large distortion with the Cu-O plane originating from size variance of the ions substituted onto the RE-site \cite{mazza_searching_2022}.

\begin{table*}[ht]
\caption{Structural information for the different compositions of (5Y)BCO. Ionic radii (I.R.) and variance are calculated using eight-fold coordinate values from the Shannon-Prewitt tables \cite{shannon1976revised}. Lattice parameters and bond angles are taken from Rietveld refinement of the synchrotron data. Error on all refined lattice parameters is smaller than the final decimal place by at least one order of magnitude. Values for unalloyed YBCO are taken from Ref. \cite{nozik1991neutron}}
\centering \includegraphics[width=7in]{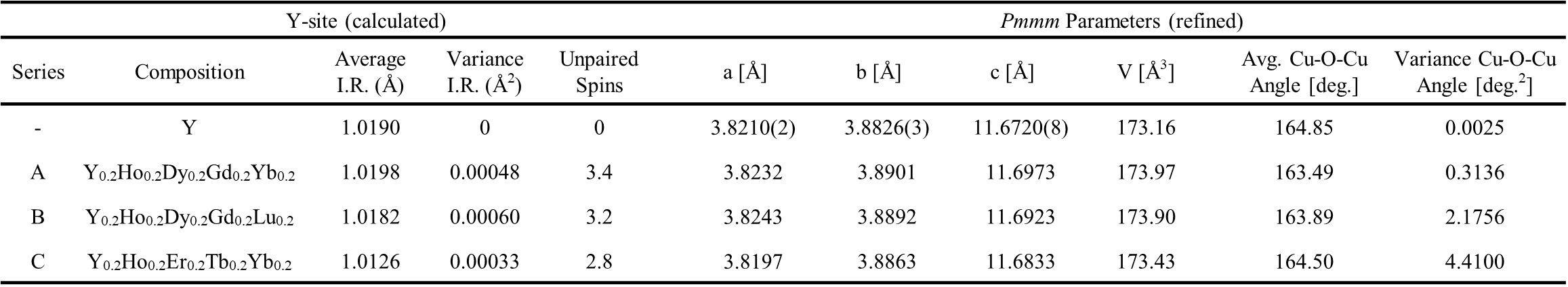}
\label{fig:T1}
\end{table*}

\begin{figure*}[ht]
\centering \includegraphics[width=7in]{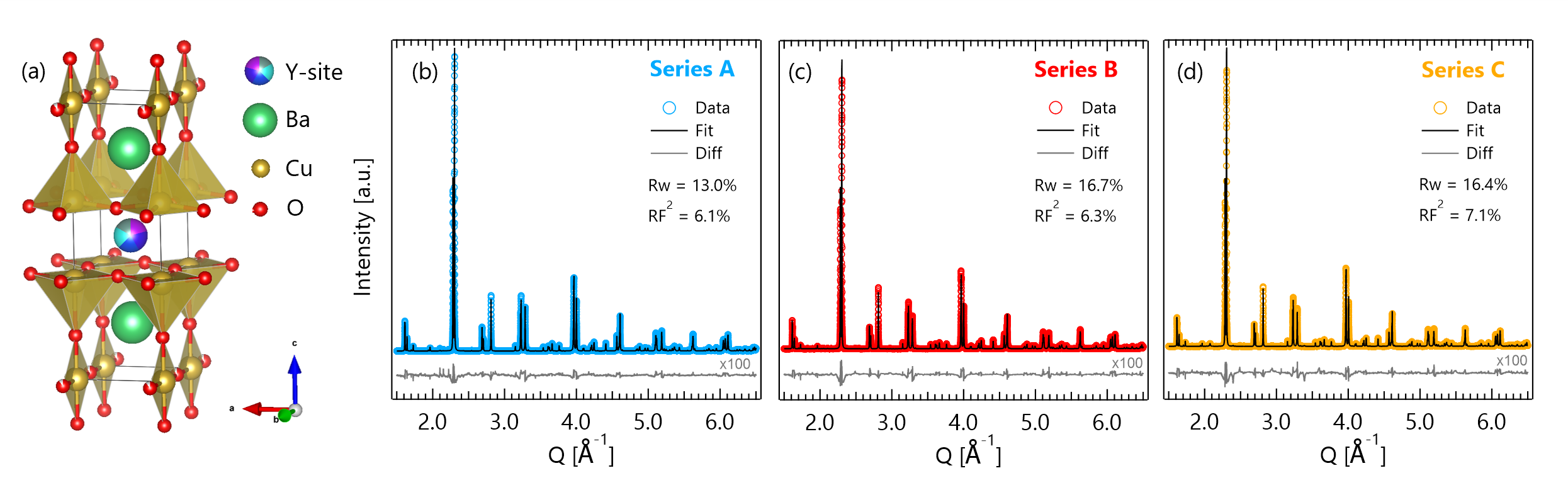}
\caption{(a) The \textit{Pmmm} crystal structure of (5Y)BCO with five elements alloyed on the Y-site. (b)-(d) Reitveld refinements of synchrotron X-ray diffraction data (APS, 11-BM) for each of the three Y-site compositions. The difference curve is weighted by the standard uncertainty defined as $(Data-Fit)/\sqrt{Data}$ and scaled by 100x for visibility. The weighted  residual, R$_W$, is a goodness-of-fit metric of the whole data set, while the unweighted phase residual, RF$^2$, is specific to the (5Y)BCO phase.}
\label{fig:sXRD}
\end{figure*}

Here, for the first time, we report the successful synthesis of compositionally-complex cuprates that exhibit high-T$_C$ superconductivity. Specifically, compositionally-complex variants of YBa$_2$Cu$_3$O$_{7-x}$ were created by alloying three different sets of trivalent (isovalent) ions onto the Y-site to create three, unique (5Y)BCO compositions with varying degrees of size and spin disorder, parameterized by the ion radii variance and average spin, respectively. High phase-purity and elemental mixing for all three (5Y)BCO compositions was confirmed by synchrotron diffraction and electron microscopy. All compositions exhibited T$_C$ values between 91 K - 92 K near optimal oxygen doping, and AC magnetometry confirmed these transitions to be superconducting. The lack of observed $T_C$ suppression in (5Y)BCO demonstrates that paring in this structure type can be remarkably robust to disorder on the Y-site. These results are discussed in terms impurity potentials and their contrast with recent experiments on (5R)CO.


Powder batches of each (5Y)BCO composition were prepared by weighing out stoichiometric amounts of metal oxide and metal carbonate precursors followed by manually grinding, room-temperature pellet pressing, and firing in a muffle furnace at dwell temperatures ranging from 900$^{\circ}$C - 1000$^{\circ}$C. Each set of Y-site ions was selected to have an average ionic radius close that of yttrium to minimize lattice strain and long-range structural distortions. Only trivalent substitutional ions were used to prevent charge doping differences between series. Doping in these samples is instead due to oxygen vacancies, the concentration of which was controlled by the quench rate during synthesis with an optimal cooling rate of 1 $^{\circ}$C/min for all series. Synthesis conditions were optimized using phase-purity and transition temperature as the primary metrics. These metrics were evaluated respectively using lab X-ray diffraction (XRD) collected on a Panalytic MPD (Cu K$\alpha$) and four-point probe measurements taken with a Quantum Design Versalab. Sample homogeneity and element clustering in sintered pellets was analyzed by collecting element-specific composition maps at different magnifications using energy dispersive X-ray spectroscopy in a FEI Nova 430 scanning electron microscope (SEM-EDS) \cite{SuppMater}. A summary of these SEM-EDS results is presented in Figs. S1-S6 showing that samples that appear phase-pure to lab-source XRD have no elemental clustering larger than 10 nm. Dozens of samples of each composition were produced during the synthesis optimization. However, the results presented in the main text were collected from a total of five samples made using recipes confirmed to reproducibly generate samples that appear phase-pure when characterized by these methods.

The room-temperature crystal structure for each composition of (5Y)BCO was determined using high-resolution synchrotron XRD measurements collected at Argonne National Lab's Advanced Photon Source 11-BM beamline and refined using the Rietveld method implemented in GSAS-II. Figure \ref{fig:sXRD} shows the synchrotron XRD data and fits corresponding to the orthorhombic \textit{Pmmm} structure exhibited by pure YBCO and expected for our (5Y)BCO samples. The \textit{Pmmm} phase fraction ranges from 99.5$\%$ in the Series A sample to a minimum of 97.5$\%$ in the Series C sample. The remaining peaks index to trace amounts of unreacted binary precursors and the intermediate ternary phase BaCuO$_2$. The difference curves show the largest remaining error in the fits corresponds to peak shape mismatch on the \textit{Pmmm} peaks. This confirms the impurities are small enough and sufficiently well fit so as not to limit the refinement accuracy of the (5Y)BCO \textit{Pmmm} phase.

The refined structural parameters for each (5Y)BCO composition are reported in Table 1 alongside a list of the substitutional Y-site elements, their average ionic radii, size disorder, and average number of unpaired spins. Size disorder ($\sigma^2$) is quantified as the variance of the ionic radii of the elements on the Y-site defined as $\sigma^2 = \sum_i x{_i}r{_i^2} - r_A^2$ where $r_A = \sum_i x{_i}r{_i}$ is the average ionic radius of the A-site cation, and $x_i$ is the fraction of the ith cation of radius $r_i$. Several notable trends are seen in these refinement results. First, a larger unit cell (0.2-0.4$\%$) is refined for each (5Y)BCO composition compared to pure YBCO. This unit cell expansion is primarily due to elongation of the c-axis (0.1-0.2$\%$), and much smaller expansions are mostly observed along the a- and b- axes. In the case of C-series, the a-axis appears to shorten slightly.  Second, the (5Y)BCO unit cell volumes and c-axis lattice constants are directly correlated with the average ionic radius of the Y-site element mixture. This is consistent with a picture that isovalent Y-site alloying creates expansive chemical pressure along the c-axis. However, unit cell and c-axis expansion still occur in C series, where the average Y-site ionic radius is significantly smaller than Y$^{3+}$, which one expects to cause c-axis compression from a simple chemical pressure model. This suggests a tensile strain contribution exists along the c-axis in these disordered (5Y)BCO compounds, which is separate from and adds to the chemical pressure expected from having different average ionic radius on the Y-site. However, in entropy-alloyed (5R)CO, c-axis contraction and smaller unit cells were observed upon adding disorder to the La-site \cite{musico_synthesis_2021}, which suggests that a pure chemical pressure model is insufficient to fully explain the effect of the disorder on structure in these materials.

The sXRD refinement results also indicate that Y-site alloying changes the Cu-O-Cu bond angles in the CuO$_2$-planes that carry supercurrent. On average, the Cu-O-Cu bond angles in the CuO$_2$-plane are smaller in (5Y)BCO than pure YBCO. However, the (5Y)BCO samples appear to have greater variance of the Cu-O-Cu bond angles (i.e., greater difference between Cu-O2-Cu vs. Cu-O3-Cu angles) than pure YBCO. Interestingly, variance in the Cu-O-Cu bond angles correlates with decreasing average ionic radii on the Y-site, rather than with the variance of the ionic radii.

\begin{figure}[h]
    \includegraphics[width=3.3in]{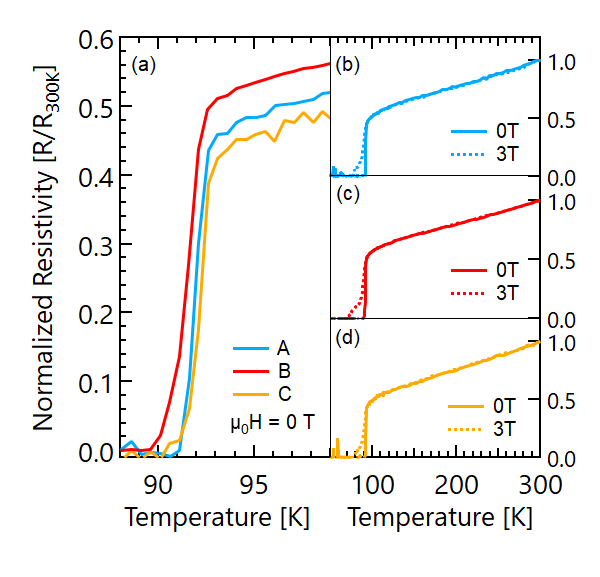}
    \caption{Electrical resistivity measurements normalized to the resistivity at 300 K. (a) Zero-field cooled (0T) normalized resistivity shown near the superconducting transition. (b)-(d) Full temperature range showing T-linear behavior above T$_C$ for Series A, B, and C, respectively.}
    \label{fig:Transport}
\end{figure}

Electrical resistivity measurements for each (5Y)BCO series, collected under 0 T and 3 T magnetic fields on pressed and sintered pellets, are shown in Figure \ref{fig:Transport} normalized to the resistivity value at 300 K. Note that all three samples presented here exhibit T-linear resistivity above T$_C$, which is known to occur primarily near optimal doping and the peak of the superconducting dome \cite{gurvitch_resistivity_1987, keimer_quantum_2015, le_tacon_strange_2021, phillips_stranger_2022}. In lieu of directly measured oxygen concentrations, which could not be accurately extracted from our sXRD refinements, the linearity of the resistivity data above T$_C$ was used to optimize cooling rate and hole doping in our samples. 

In the near-optimally doped samples shown in Fig. \ref{fig:Transport}, all (5Y)BCO compositions show a sharp drop to zero resistance above 90 K. T$_C$ values for the three samples range from 91.5 K - 92.5 K, calculated as the inflection point of the normalized resistance change. Application of a 3 T magnetic field suppresses and broadens the resistivity transition. This behavior is expected and has been reported previously for the parent YBCO, and is associated with a multi-critical point where the superconducting transition goes from first-order to second-order \cite{langan_phase_1998}. The T$_C$ values for samples \textit{within} a series were observed to vary on the order of 1 K from sample to sample, despite nominally identical synthesis procedures \cite{SuppMater}. Therefore, no conclusions should be drawn from the order of T$_C$ values presented in Fig. \ref{fig:Transport} with respect to their chemical or structural differences, as T$_C$ ordering is not repeatable. What is repeatable is that for an optimized synthesis procedure, T$_C$ was always observed to be $>$91 K, indicating that the spin and lattice disorder added to the system do not greatly disrupt the pairing interaction.

\begin{figure}[h]
    \includegraphics[width=3.5in]{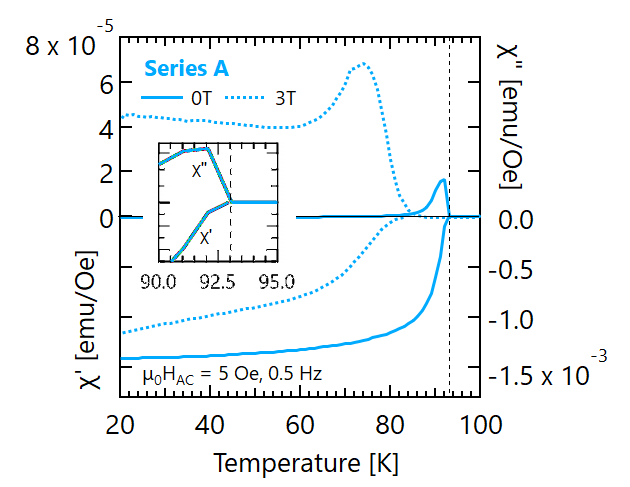}
    \caption{Zero-field cooled (0T) and field-cooled (3T) AC magnetic susceptibility measurements from a Series A sample. Inset shows a close up of the zero-field transition.}
    \label{fig:AC_SQUID}
\end{figure}

The superconducting nature of the resistivity transition was verified using AC susceptibility measurements, collected in a Quantum Design MPMS3 Superconducting Quantum Interference Device (SQUID) magnetometer. Figure \ref{fig:AC_SQUID} shows AC susceptibility data collected on the same Series A sample shown in Fig. \ref{fig:Transport}. $\chi'$ and $\chi''$ are the real and imaginary part of susceptibility of the material, respectively, and the divergence between the two marks the superconducting transition temperature T$_C$. For Series A under zero-field cooling, this divergence occurs just above 92.5 K, which is in perfect agreement with the start of the resistance downtown in Fig. \ref{fig:Transport}. The peak in $\chi''$ occurs near 92 K, which aligns with the inflection of the resistance transition in this sample. The downturn in $\chi'$ confirms the diamagnetic Meissner state below T$_C$. Additionally, the shape of the $\chi''$ peak gives information about flux pinning in these samples. Under the zero field, the $\chi''$ peak is narrow suggesting strong, uniform flux pinning sites in the Meissner state. Under a 3 T applied field, the $\chi''$ peak broadens and shifts lower in temperature, reflecting a weaker and wider range of pinning interactions in the Type-II vortex state.

The key result from this study is the unambiguous observation of high-temperature superconductivity in a series of compositionally-complex and disordered cuprates with the YBCO structure. This result and the (5Y)BCO copmounds introduced here are intriguing for at least three reasons: (1) the lack of $T_C$ suppression, (2) the contrasting superconducting behavior with doped (5R)CO films, and (3) the potential for these systems to disentangle the effects of spin and lattice disorder and identify the dominant pairing interaction in high-temperature d-wave superconductors.

Considering first the minimal $T_C$ suppression observed in (5Y)BCO, we leverage impurity potentials as a framework for comparing the degree to which different impurities are expected to disrupt pairing and suppress superconductivity. The strength of impurity potentials in cuprates depends both on the distance of the impurity from the superconducting plane and its charge difference with the parent ion on the impurity site \cite{graser_t_2007, ozdemir_effect_2022}. Because only isovalent impurities were used in this study, the (5Y)BCO lattice ``sees" no apparent charge disorder, and thus impurity potentials are generally expected to be weak in this compounds. This implies minimal pairing disruption should be expected here, and agrees well with the near-maximum $T_C$ values we observed. Further support of this picture is seen by comparison with prior studies where only a single element was alloyed onto the Y-site o YBCO \cite{devi_enhanced_2000, macmanus-driscoll_rare_2005,haugan_microstructural_2008,wee_formation_2011}, which show similarly little $T_C$ suppression. Thus the lack of $T_C$ suppression observed in (5Y)BCO is consistent with both theory and prior experimental work on chemically-simpler YBCO alloys.

Considering now the contrasting results reported for (5Y)BCO and (5R)CO, we first note that in both systems the disordered site is plaquette-centered above the CuO$_2$ square lattice and roughly the same distance from the CuO$_2$ plane. Specifically, in YBCO, the Y-site sits roughly 3.2 $\AA$ from Cu and 2.4 $\AA$ from O, while in the T'-(5R)CO structure, the RE-site sits slightly further away at 3.2-3.3 $\AA$ from Cu and 2.7 $\AA$ from O \cite{nozik1991neutron,tuilier1992exafs}. These small differences between the Y-site and La-site position with respect to its neighboring CuO$_2$ plane suggest these materials should experience similar impurity potentials for \textit{isovalent} disorder on these sites. However, charge doping (5R)CO requires use on divalent or tetravalent ions on the RE-site, which creates a charge difference on the RE-site relative to trivalent La and therefore significantly increases the impurity potential in the (5R)CO system. Thus, larger impurity potentials are expected for the doped (5R)CO than (5Y)BCO with oxygen doping, which could explain the lack of superconductivity observed in those samples.

Alternatively, Mazza et al. proposed that the large cation-size disorder on the RE-site was the reason their (5R)CO samples did not superconduct \cite{mazza_searching_2022}.  Using the Shannon-Prewitt ionic radii \cite{shannon1976revised}, the reported (5R)CO samples have calculated size disorder parameters of 0.0025 and 0.0030 $\AA^2$ for the electron- and hole-doped samples, respectively. By comparison, the (5Y)BCO samples reported here have size disorder parameters that vary from 0.0003-0.0006 $\AA^2$, approximately one order of magnitude smaller than the (5R)CO samples. This analysis is consistent the authors' hypothesis that large size disorder suppresses superconductivity in their (5R)CO samples. However, additional studies are required to determine whether differences in size disorder, impurity potential, or some combination is responsible for suppressing superconductivity in (5R)CO samples to-date.  

Last but not least, we highlight the potential of compositionally-complex cuprates to disentangle spin and lattice contributions to pairing in high-$T_C$ superconductors, the precise origin of which remains a major open question in condensed matter physics. Broadly speaking there exist two leading explanations for pairing in cuprates. The first is that pairing is mediated by antiferromagnetic spin fluctuations, which is supported by inelastic scattering experiments \cite{wakimoto_direct_2004,le_tacon_intense_2011,scalapino_common_2012,proust_remarkable_2019}. The second is that pairing is mediated by strong electron-phonon coupling, which is supported by photoemission and tunneling spectroscopy measurements \cite{lanzara_evidence_2001,iwasawa_isotopic_2008, lee_interplay_2006}. To determine which of these models is correct, one would like to isolate the spin and lattice degrees of freedom and observing their independent effects on $T_C$. However, when alloying with a single element (e.g., substituting Gd for Y in YBCO), changes in average ionic radius (lattice distortion) cannot be separated from changes in the average magnetic moment (spin distortion). As shown in Table 1, the chemical flexibility of compositionally-complex cuprates offers an ability to tune spin and lattice disorder in a decoupled and non-monotonic fashion. By extending this paradigm to compositionally-complex cuprates where disorder is more strongly or completely isolated to either the lattice or spin degree of freedom, we believe this materials class opens up a new avenue to probe pairing in high-$T_C$ superconductors.


In conclusion, this work reports high-temperature superconductivity in a family of compositionally-complex cuprate superconductors, and demonstrates the possibility of decoupling size and spin disorder for the purpose of interrogating pairing interactions in these materials. Three compositions of (5Y)BCO were made with high phase-purity and $T_C$ values greater than 91 K. The negligible $T_C$ reduction relative to undisordered YBCO is consistent with prior work on simpler YBCO alloys and a theoretical framework of weak impurity potentials in isovalently alloyed systems. The contrasting behavior of (5Y)BCO with recent reports of compositionally-complex (5R)CO highlights the sensitive nature of d-wave superconductivity to different types of disorder, and provides clear opportunities for future studies.

\section{Acknowledgements}
We thank T.Z. Ward and P.J. Hirschfeld for helpful conversations designing this project and interpreting the results. This work was supported by UF Research. N.M.C. and E.W. were supported by the NSF REU program (DMR-9820518).

\section{Author Contributions}
A.R., N.A., and R.F.N. conceptualized the project and designed the experiments. Samples were synthesized and characterized by A.R., N.M.C., S.D., R.N., E.W., M.W., C.G., and K.N. Synchrotron measurements were prepared by A.R. and K.S., and analyzed by A.R. and R.F.N. N.A. collected the magnetometry. A.R. and R.F.N. wrote the first manuscript draft. All authors contributed to the final version.

\bibliography{5YBCO}

\begin{thebibliography}{41}%
\makeatletter
\providecommand \@ifxundefined [1]{%
 \@ifx{#1\undefined}
}%
\providecommand \@ifnum [1]{%
 \ifnum #1\expandafter \@firstoftwo
 \else \expandafter \@secondoftwo
 \fi
}%
\providecommand \@ifx [1]{%
 \ifx #1\expandafter \@firstoftwo
 \else \expandafter \@secondoftwo
 \fi
}%
\providecommand \natexlab [1]{#1}%
\providecommand \enquote  [1]{``#1''}%
\providecommand \bibnamefont  [1]{#1}%
\providecommand \bibfnamefont [1]{#1}%
\providecommand \citenamefont [1]{#1}%
\providecommand \href@noop [0]{\@secondoftwo}%
\providecommand \href [0]{\begingroup \@sanitize@url \@href}%
\providecommand \@href[1]{\@@startlink{#1}\@@href}%
\providecommand \@@href[1]{\endgroup#1\@@endlink}%
\providecommand \@sanitize@url [0]{\catcode `\\12\catcode `\$12\catcode `\&12\catcode `\#12\catcode `\^12\catcode `\_12\catcode `\%12\relax}%
\providecommand \@@startlink[1]{}%
\providecommand \@@endlink[0]{}%
\providecommand \url  [0]{\begingroup\@sanitize@url \@url }%
\providecommand \@url [1]{\endgroup\@href {#1}{\urlprefix }}%
\providecommand \urlprefix  [0]{URL }%
\providecommand \Eprint [0]{\href }%
\providecommand \doibase [0]{http://dx.doi.org/}%
\providecommand \selectlanguage [0]{\@gobble}%
\providecommand \bibinfo  [0]{\@secondoftwo}%
\providecommand \bibfield  [0]{\@secondoftwo}%
\providecommand \translation [1]{[#1]}%
\providecommand \BibitemOpen [0]{}%
\providecommand \bibitemStop [0]{}%
\providecommand \bibitemNoStop [0]{.\EOS\space}%
\providecommand \EOS [0]{\spacefactor3000\relax}%
\providecommand \BibitemShut  [1]{\csname bibitem#1\endcsname}%
\let\auto@bib@innerbib\@empty
\bibitem [{\citenamefont {Anderson}(1959)}]{anderson_theory_1959}%
  \BibitemOpen
  \bibfield  {author} {\bibinfo {author} {\bibfnamefont {P.~W.}\ \bibnamefont {Anderson}},\ }\href {\doibase https://doi.org/10.1016/0022-3697(59)90036-8} {\bibfield  {journal} {\bibinfo  {journal} {J. Phys. Chem. Solids}\ }\textbf {\bibinfo {volume} {11}},\ \bibinfo {pages} {26} (\bibinfo {year} {1959})}\BibitemShut {NoStop}%
\bibitem [{\citenamefont {Fisher}(1990)}]{fisher_quantum_1990}%
  \BibitemOpen
  \bibfield  {author} {\bibinfo {author} {\bibfnamefont {M.~P.~A.}\ \bibnamefont {Fisher}},\ }\href {\doibase 10.1103/PhysRevLett.65.923} {\bibfield  {journal} {\bibinfo  {journal} {Phys. Rev. Lett.}\ }\textbf {\bibinfo {volume} {65}},\ \bibinfo {pages} {923} (\bibinfo {year} {1990})}\BibitemShut {NoStop}%
\bibitem [{\citenamefont {Pan}\ \emph {et~al.}(2001)\citenamefont {Pan}, \citenamefont {O'Neal}, \citenamefont {Badzey}, \citenamefont {Chamon}, \citenamefont {Ding}, \citenamefont {Engelbrecht}, \citenamefont {Wang}, \citenamefont {Eisaki}, \citenamefont {Uchida},\ and\ \citenamefont {Gupta}}]{pan_microscopic_2001}%
  \BibitemOpen
  \bibfield  {author} {\bibinfo {author} {\bibfnamefont {S.~H.}\ \bibnamefont {Pan}}, \bibinfo {author} {\bibfnamefont {J.~P.}\ \bibnamefont {O'Neal}}, \bibinfo {author} {\bibfnamefont {R.~L.}\ \bibnamefont {Badzey}}, \bibinfo {author} {\bibfnamefont {C.}~\bibnamefont {Chamon}}, \bibinfo {author} {\bibfnamefont {H.}~\bibnamefont {Ding}}, \bibinfo {author} {\bibfnamefont {J.~R.}\ \bibnamefont {Engelbrecht}}, \bibinfo {author} {\bibfnamefont {Z.}~\bibnamefont {Wang}}, \bibinfo {author} {\bibfnamefont {H.}~\bibnamefont {Eisaki}}, \bibinfo {author} {\bibfnamefont {S.}~\bibnamefont {Uchida}}, \ and\ \bibinfo {author} {\bibfnamefont {A.~K.}\ \bibnamefont {Gupta}},\ }\href@noop {} {\bibfield  {journal} {\bibinfo  {journal} {Nature}\ }\textbf {\bibinfo {volume} {413}} (\bibinfo {year} {2001})}\BibitemShut {NoStop}%
\bibitem [{\citenamefont {Dubi}\ \emph {et~al.}(2007)\citenamefont {Dubi}, \citenamefont {Meir},\ and\ \citenamefont {Avishai}}]{dubi_nature_2007}%
  \BibitemOpen
  \bibfield  {author} {\bibinfo {author} {\bibfnamefont {Y.}~\bibnamefont {Dubi}}, \bibinfo {author} {\bibfnamefont {Y.}~\bibnamefont {Meir}}, \ and\ \bibinfo {author} {\bibfnamefont {Y.}~\bibnamefont {Avishai}},\ }\href {\doibase 10.1038/nature06180} {\bibfield  {journal} {\bibinfo  {journal} {Nature}\ }\textbf {\bibinfo {volume} {449}},\ \bibinfo {pages} {876} (\bibinfo {year} {2007})}\BibitemShut {NoStop}%
\bibitem [{\citenamefont {Sacépé}\ \emph {et~al.}(2011)\citenamefont {Sacépé}, \citenamefont {Dubouchet}, \citenamefont {Chapelier}, \citenamefont {Sanquer}, \citenamefont {Ovadia}, \citenamefont {Shahar}, \citenamefont {Feigel’man},\ and\ \citenamefont {Ioffe}}]{sacepe_localization_2011}%
  \BibitemOpen
  \bibfield  {author} {\bibinfo {author} {\bibfnamefont {B.}~\bibnamefont {Sacépé}}, \bibinfo {author} {\bibfnamefont {T.}~\bibnamefont {Dubouchet}}, \bibinfo {author} {\bibfnamefont {C.}~\bibnamefont {Chapelier}}, \bibinfo {author} {\bibfnamefont {M.}~\bibnamefont {Sanquer}}, \bibinfo {author} {\bibfnamefont {M.}~\bibnamefont {Ovadia}}, \bibinfo {author} {\bibfnamefont {D.}~\bibnamefont {Shahar}}, \bibinfo {author} {\bibfnamefont {M.}~\bibnamefont {Feigel’man}}, \ and\ \bibinfo {author} {\bibfnamefont {L.}~\bibnamefont {Ioffe}},\ }\href {\doibase 10.1038/nphys1892} {\bibfield  {journal} {\bibinfo  {journal} {Nat. Phys.}\ }\textbf {\bibinfo {volume} {7}},\ \bibinfo {pages} {239} (\bibinfo {year} {2011})}\BibitemShut {NoStop}%
\bibitem [{\citenamefont {Mizuguchi}\ \emph {et~al.}(2023)\citenamefont {Mizuguchi}, \citenamefont {Usui}, \citenamefont {Kurita}, \citenamefont {Takae}, \citenamefont {Kasem}, \citenamefont {Matsumoto}, \citenamefont {Yamane}, \citenamefont {Takano}, \citenamefont {Nakahira}, \citenamefont {Yamashita}, \citenamefont {Goto}, \citenamefont {Miura},\ and\ \citenamefont {Moriyoshi}}]{mizuguchi_glassy_2023}%
  \BibitemOpen
  \bibfield  {author} {\bibinfo {author} {\bibfnamefont {Y.}~\bibnamefont {Mizuguchi}}, \bibinfo {author} {\bibfnamefont {H.}~\bibnamefont {Usui}}, \bibinfo {author} {\bibfnamefont {R.}~\bibnamefont {Kurita}}, \bibinfo {author} {\bibfnamefont {K.}~\bibnamefont {Takae}}, \bibinfo {author} {\bibfnamefont {M.~R.}\ \bibnamefont {Kasem}}, \bibinfo {author} {\bibfnamefont {R.}~\bibnamefont {Matsumoto}}, \bibinfo {author} {\bibfnamefont {K.}~\bibnamefont {Yamane}}, \bibinfo {author} {\bibfnamefont {Y.}~\bibnamefont {Takano}}, \bibinfo {author} {\bibfnamefont {Y.}~\bibnamefont {Nakahira}}, \bibinfo {author} {\bibfnamefont {A.}~\bibnamefont {Yamashita}}, \bibinfo {author} {\bibfnamefont {Y.}~\bibnamefont {Goto}}, \bibinfo {author} {\bibfnamefont {A.}~\bibnamefont {Miura}}, \ and\ \bibinfo {author} {\bibfnamefont {C.}~\bibnamefont {Moriyoshi}},\ }\href {\doibase 10.1016/j.mtphys.2023.101019} {\bibfield  {journal} {\bibinfo  {journal} {Mater. Today Phys.}\ }\textbf {\bibinfo {volume} {32}},\ \bibinfo {pages} {101019}
  (\bibinfo {year} {2023})}\BibitemShut {NoStop}%
\bibitem [{\citenamefont {Ma}\ and\ \citenamefont {Lee}(1985)}]{ma_localized_1985}%
  \BibitemOpen
  \bibfield  {author} {\bibinfo {author} {\bibfnamefont {M.}~\bibnamefont {Ma}}\ and\ \bibinfo {author} {\bibfnamefont {P.~A.}\ \bibnamefont {Lee}},\ }\href {\doibase 10.1103/PhysRevB.32.5658} {\bibfield  {journal} {\bibinfo  {journal} {Phys. Rev. B}\ }\textbf {\bibinfo {volume} {32}},\ \bibinfo {pages} {5658} (\bibinfo {year} {1985})}\BibitemShut {NoStop}%
\bibitem [{\citenamefont {Hebard}\ and\ \citenamefont {Paalanen}(1990)}]{hebard_magnetic-field-tuned_1990}%
  \BibitemOpen
  \bibfield  {author} {\bibinfo {author} {\bibfnamefont {A.~F.}\ \bibnamefont {Hebard}}\ and\ \bibinfo {author} {\bibfnamefont {M.~A.}\ \bibnamefont {Paalanen}},\ }\href {\doibase 10.1103/PhysRevLett.65.927} {\bibfield  {journal} {\bibinfo  {journal} {Phys. Rev. Lett.}\ }\textbf {\bibinfo {volume} {65}},\ \bibinfo {pages} {927} (\bibinfo {year} {1990})}\BibitemShut {NoStop}%
\bibitem [{\citenamefont {Abrikosov}\ and\ \citenamefont {Gor'kov}(1960)}]{abrikosov1960contribution}%
  \BibitemOpen
  \bibfield  {author} {\bibinfo {author} {\bibfnamefont {A.~A.}\ \bibnamefont {Abrikosov}}\ and\ \bibinfo {author} {\bibfnamefont {L.~P.}\ \bibnamefont {Gor'kov}},\ }\href@noop {} {\bibfield  {journal} {\bibinfo  {journal} {Zhur. Eksptl'. i Teoret. Fiz.}\ }\textbf {\bibinfo {volume} {39}} (\bibinfo {year} {1960})}\BibitemShut {NoStop}%
\bibitem [{\citenamefont {Tolpygo}\ \emph {et~al.}(1996)\citenamefont {Tolpygo}, \citenamefont {Lin}, \citenamefont {Gurvitch}, \citenamefont {Hou},\ and\ \citenamefont {Phillips}}]{tolpygo_universal_1996}%
  \BibitemOpen
  \bibfield  {author} {\bibinfo {author} {\bibfnamefont {S.~K.}\ \bibnamefont {Tolpygo}}, \bibinfo {author} {\bibfnamefont {J.-Y.}\ \bibnamefont {Lin}}, \bibinfo {author} {\bibfnamefont {M.}~\bibnamefont {Gurvitch}}, \bibinfo {author} {\bibfnamefont {S.~Y.}\ \bibnamefont {Hou}}, \ and\ \bibinfo {author} {\bibfnamefont {J.~M.}\ \bibnamefont {Phillips}},\ }\href {\doibase 10.1103/PhysRevB.53.12454} {\bibfield  {journal} {\bibinfo  {journal} {Phys. Rev. B}\ }\textbf {\bibinfo {volume} {53}},\ \bibinfo {pages} {12454} (\bibinfo {year} {1996})}\BibitemShut {NoStop}%
\bibitem [{\citenamefont {Fujita}\ \emph {et~al.}(2005)\citenamefont {Fujita}, \citenamefont {Noda}, \citenamefont {Kojima}, \citenamefont {Eisaki},\ and\ \citenamefont {Uchida}}]{fujita_effect_2005}%
  \BibitemOpen
  \bibfield  {author} {\bibinfo {author} {\bibfnamefont {K.}~\bibnamefont {Fujita}}, \bibinfo {author} {\bibfnamefont {T.}~\bibnamefont {Noda}}, \bibinfo {author} {\bibfnamefont {K.~M.}\ \bibnamefont {Kojima}}, \bibinfo {author} {\bibfnamefont {H.}~\bibnamefont {Eisaki}}, \ and\ \bibinfo {author} {\bibfnamefont {S.}~\bibnamefont {Uchida}},\ }\href {\doibase 10.1103/PhysRevLett.95.097006} {\bibfield  {journal} {\bibinfo  {journal} {Phys. Rev. Lett.}\ }\textbf {\bibinfo {volume} {95}},\ \bibinfo {pages} {097006} (\bibinfo {year} {2005})}\BibitemShut {NoStop}%
\bibitem [{\citenamefont {Sun}\ and\ \citenamefont {Cava}(2019)}]{sun_high-entropy_2019}%
  \BibitemOpen
  \bibfield  {author} {\bibinfo {author} {\bibfnamefont {L.}~\bibnamefont {Sun}}\ and\ \bibinfo {author} {\bibfnamefont {R.~J.}\ \bibnamefont {Cava}},\ }\href {\doibase 10.1103/PhysRevMaterials.3.090301} {\bibfield  {journal} {\bibinfo  {journal} {Phys. Rev. Mater.}\ }\textbf {\bibinfo {volume} {3}},\ \bibinfo {pages} {090301} (\bibinfo {year} {2019})}\BibitemShut {NoStop}%
\bibitem [{\citenamefont {Koželj}\ \emph {et~al.}(2014)\citenamefont {Koželj}, \citenamefont {Vrtnik}, \citenamefont {Jelen}, \citenamefont {Jazbec}, \citenamefont {Jagličić}, \citenamefont {Maiti}, \citenamefont {Feuerbacher}, \citenamefont {Steurer},\ and\ \citenamefont {Dolinšek}}]{kozelj_discovery_2014}%
  \BibitemOpen
  \bibfield  {author} {\bibinfo {author} {\bibfnamefont {P.}~\bibnamefont {Koželj}}, \bibinfo {author} {\bibfnamefont {S.}~\bibnamefont {Vrtnik}}, \bibinfo {author} {\bibfnamefont {A.}~\bibnamefont {Jelen}}, \bibinfo {author} {\bibfnamefont {S.}~\bibnamefont {Jazbec}}, \bibinfo {author} {\bibfnamefont {Z.}~\bibnamefont {Jagličić}}, \bibinfo {author} {\bibfnamefont {S.}~\bibnamefont {Maiti}}, \bibinfo {author} {\bibfnamefont {M.}~\bibnamefont {Feuerbacher}}, \bibinfo {author} {\bibfnamefont {W.}~\bibnamefont {Steurer}}, \ and\ \bibinfo {author} {\bibfnamefont {J.}~\bibnamefont {Dolinšek}},\ }\href {\doibase 10.1103/PhysRevLett.113.107001} {\bibfield  {journal} {\bibinfo  {journal} {Phys. Rev. Lett.}\ }\textbf {\bibinfo {volume} {113}},\ \bibinfo {pages} {107001} (\bibinfo {year} {2014})}\BibitemShut {NoStop}%
\bibitem [{\citenamefont {Von~Rohr}\ \emph {et~al.}(2016)\citenamefont {Von~Rohr}, \citenamefont {Winiarski}, \citenamefont {Tao}, \citenamefont {Klimczuk},\ and\ \citenamefont {Cava}}]{von_rohr_effect_2016}%
  \BibitemOpen
  \bibfield  {author} {\bibinfo {author} {\bibfnamefont {F.}~\bibnamefont {Von~Rohr}}, \bibinfo {author} {\bibfnamefont {M.~J.}\ \bibnamefont {Winiarski}}, \bibinfo {author} {\bibfnamefont {J.}~\bibnamefont {Tao}}, \bibinfo {author} {\bibfnamefont {T.}~\bibnamefont {Klimczuk}}, \ and\ \bibinfo {author} {\bibfnamefont {R.~J.}\ \bibnamefont {Cava}},\ }\href {\doibase 10.1073/pnas.1615926113} {\bibfield  {journal} {\bibinfo  {journal} {Proc. Natl. Acad. Sci.}\ }\textbf {\bibinfo {volume} {113}} (\bibinfo {year} {2016}),\ 10.1073/pnas.1615926113}\BibitemShut {NoStop}%
\bibitem [{\citenamefont {Hirai}\ \emph {et~al.}(2023)\citenamefont {Hirai}, \citenamefont {Uematsu}, \citenamefont {Saitoh}, \citenamefont {Katayama},\ and\ \citenamefont {Takenaka}}]{hirai_superconductivity_2023}%
  \BibitemOpen
  \bibfield  {author} {\bibinfo {author} {\bibfnamefont {D.}~\bibnamefont {Hirai}}, \bibinfo {author} {\bibfnamefont {N.}~\bibnamefont {Uematsu}}, \bibinfo {author} {\bibfnamefont {K.}~\bibnamefont {Saitoh}}, \bibinfo {author} {\bibfnamefont {N.}~\bibnamefont {Katayama}}, \ and\ \bibinfo {author} {\bibfnamefont {K.}~\bibnamefont {Takenaka}},\ }\href {\doibase 10.1021/acs.inorgchem.3c01364} {\bibfield  {journal} {\bibinfo  {journal} {Inorg. Chem.}\ ,\ \bibinfo {pages} {acs.inorgchem.3c01364}} (\bibinfo {year} {2023})}\BibitemShut {NoStop}%
\bibitem [{\citenamefont {Musicó}\ \emph {et~al.}(2021)\citenamefont {Musicó}, \citenamefont {Wright}, \citenamefont {Delzer}, \citenamefont {Ward}, \citenamefont {Rawn}, \citenamefont {Mandrus},\ and\ \citenamefont {Keppens}}]{musico_synthesis_2021}%
  \BibitemOpen
  \bibfield  {author} {\bibinfo {author} {\bibfnamefont {B.~L.}\ \bibnamefont {Musicó}}, \bibinfo {author} {\bibfnamefont {Q.}~\bibnamefont {Wright}}, \bibinfo {author} {\bibfnamefont {C.}~\bibnamefont {Delzer}}, \bibinfo {author} {\bibfnamefont {T.~Z.}\ \bibnamefont {Ward}}, \bibinfo {author} {\bibfnamefont {C.~J.}\ \bibnamefont {Rawn}}, \bibinfo {author} {\bibfnamefont {D.~G.}\ \bibnamefont {Mandrus}}, \ and\ \bibinfo {author} {\bibfnamefont {V.}~\bibnamefont {Keppens}},\ }\href {\doibase 10.1111/jace.17750} {\bibfield  {journal} {\bibinfo  {journal} {J. Am. Ceram. Soc.}\ }\textbf {\bibinfo {volume} {104}},\ \bibinfo {pages} {3750} (\bibinfo {year} {2021})}\BibitemShut {NoStop}%
\bibitem [{\citenamefont {Mazza}\ \emph {et~al.}(2022)\citenamefont {Mazza}, \citenamefont {Gao}, \citenamefont {Rossi}, \citenamefont {Musico}, \citenamefont {Valentine}, \citenamefont {Kennedy}, \citenamefont {Zhang}, \citenamefont {Lapano}, \citenamefont {Keppens}, \citenamefont {Moore}, \citenamefont {Brahlek}, \citenamefont {Rost},\ and\ \citenamefont {Ward}}]{mazza_searching_2022}%
  \BibitemOpen
  \bibfield  {author} {\bibinfo {author} {\bibfnamefont {A.~R.}\ \bibnamefont {Mazza}}, \bibinfo {author} {\bibfnamefont {X.}~\bibnamefont {Gao}}, \bibinfo {author} {\bibfnamefont {D.~J.}\ \bibnamefont {Rossi}}, \bibinfo {author} {\bibfnamefont {B.~L.}\ \bibnamefont {Musico}}, \bibinfo {author} {\bibfnamefont {T.~W.}\ \bibnamefont {Valentine}}, \bibinfo {author} {\bibfnamefont {Z.}~\bibnamefont {Kennedy}}, \bibinfo {author} {\bibfnamefont {J.}~\bibnamefont {Zhang}}, \bibinfo {author} {\bibfnamefont {J.}~\bibnamefont {Lapano}}, \bibinfo {author} {\bibfnamefont {V.}~\bibnamefont {Keppens}}, \bibinfo {author} {\bibfnamefont {R.~G.}\ \bibnamefont {Moore}}, \bibinfo {author} {\bibfnamefont {M.}~\bibnamefont {Brahlek}}, \bibinfo {author} {\bibfnamefont {C.~M.}\ \bibnamefont {Rost}}, \ and\ \bibinfo {author} {\bibfnamefont {T.~Z.}\ \bibnamefont {Ward}},\ }\href {\doibase 10.1116/6.0001441} {\bibfield  {journal} {\bibinfo  {journal} {J. Vac. Sci. \& Technol. A}\ }\textbf {\bibinfo {volume} {40}},\ \bibinfo {pages}
  {013404} (\bibinfo {year} {2022})}\BibitemShut {NoStop}%
\bibitem [{\citenamefont {Rost}\ \emph {et~al.}(2015)\citenamefont {Rost}, \citenamefont {Sachet}, \citenamefont {Borman}, \citenamefont {Moballegh}, \citenamefont {Dickey}, \citenamefont {Hou}, \citenamefont {Jones}, \citenamefont {Curtarolo},\ and\ \citenamefont {Maria}}]{rost_entropy-stabilized_2015}%
  \BibitemOpen
  \bibfield  {author} {\bibinfo {author} {\bibfnamefont {C.~M.}\ \bibnamefont {Rost}}, \bibinfo {author} {\bibfnamefont {E.}~\bibnamefont {Sachet}}, \bibinfo {author} {\bibfnamefont {T.}~\bibnamefont {Borman}}, \bibinfo {author} {\bibfnamefont {A.}~\bibnamefont {Moballegh}}, \bibinfo {author} {\bibfnamefont {E.~C.}\ \bibnamefont {Dickey}}, \bibinfo {author} {\bibfnamefont {D.}~\bibnamefont {Hou}}, \bibinfo {author} {\bibfnamefont {J.~L.}\ \bibnamefont {Jones}}, \bibinfo {author} {\bibfnamefont {S.}~\bibnamefont {Curtarolo}}, \ and\ \bibinfo {author} {\bibfnamefont {J.-P.}\ \bibnamefont {Maria}},\ }\href {\doibase 10.1038/ncomms9485} {\bibfield  {journal} {\bibinfo  {journal} {Nat. Commun.}\ }\textbf {\bibinfo {volume} {6}},\ \bibinfo {pages} {8485} (\bibinfo {year} {2015})}\BibitemShut {NoStop}%
\bibitem [{\citenamefont {Zhang}\ \emph {et~al.}(2020)\citenamefont {Zhang}, \citenamefont {Mazza}, \citenamefont {Skoropata}, \citenamefont {Mukherjee}, \citenamefont {Musico}, \citenamefont {Zhang}, \citenamefont {Keppens}, \citenamefont {Zhang}, \citenamefont {Kisslinger}, \citenamefont {Stavitski}, \citenamefont {Brahlek}, \citenamefont {Freeland}, \citenamefont {Lu},\ and\ \citenamefont {Ward}}]{zhang_applying_2020}%
  \BibitemOpen
  \bibfield  {author} {\bibinfo {author} {\bibfnamefont {W.}~\bibnamefont {Zhang}}, \bibinfo {author} {\bibfnamefont {A.~R.}\ \bibnamefont {Mazza}}, \bibinfo {author} {\bibfnamefont {E.}~\bibnamefont {Skoropata}}, \bibinfo {author} {\bibfnamefont {D.}~\bibnamefont {Mukherjee}}, \bibinfo {author} {\bibfnamefont {B.}~\bibnamefont {Musico}}, \bibinfo {author} {\bibfnamefont {J.}~\bibnamefont {Zhang}}, \bibinfo {author} {\bibfnamefont {V.~M.}\ \bibnamefont {Keppens}}, \bibinfo {author} {\bibfnamefont {L.}~\bibnamefont {Zhang}}, \bibinfo {author} {\bibfnamefont {K.}~\bibnamefont {Kisslinger}}, \bibinfo {author} {\bibfnamefont {E.}~\bibnamefont {Stavitski}}, \bibinfo {author} {\bibfnamefont {M.}~\bibnamefont {Brahlek}}, \bibinfo {author} {\bibfnamefont {J.~W.}\ \bibnamefont {Freeland}}, \bibinfo {author} {\bibfnamefont {P.}~\bibnamefont {Lu}}, \ and\ \bibinfo {author} {\bibfnamefont {T.~Z.}\ \bibnamefont {Ward}},\ }\href {\doibase 10.1021/acsnano.0c04487} {\bibfield  {journal} {\bibinfo  {journal} {ACS Nano}\
  }\textbf {\bibinfo {volume} {14}},\ \bibinfo {pages} {13030} (\bibinfo {year} {2020})}\BibitemShut {NoStop}%
\bibitem [{\citenamefont {Shannon}(1976)}]{shannon1976revised}%
  \BibitemOpen
  \bibfield  {author} {\bibinfo {author} {\bibfnamefont {R.~D.}\ \bibnamefont {Shannon}},\ }\href@noop {} {\bibfield  {journal} {\bibinfo  {journal} {Acta Crystallogr. A}\ }\textbf {\bibinfo {volume} {32}},\ \bibinfo {pages} {751} (\bibinfo {year} {1976})}\BibitemShut {NoStop}%
\bibitem [{\citenamefont {Nozik}\ \emph {et~al.}(1991)\citenamefont {Nozik}, \citenamefont {Kuklina}, \citenamefont {Schuster}, \citenamefont {Weiss},\ and\ \citenamefont {Matts}}]{nozik1991neutron}%
  \BibitemOpen
  \bibfield  {author} {\bibinfo {author} {\bibfnamefont {Y.~Z.}\ \bibnamefont {Nozik}}, \bibinfo {author} {\bibfnamefont {E.}~\bibnamefont {Kuklina}}, \bibinfo {author} {\bibfnamefont {G.}~\bibnamefont {Schuster}}, \bibinfo {author} {\bibfnamefont {L.}~\bibnamefont {Weiss}}, \ and\ \bibinfo {author} {\bibfnamefont {V.}~\bibnamefont {Matts}},\ }\href@noop {} {\bibfield  {journal} {\bibinfo  {journal} {Sov. Phys. Crystallogr.}\ }\textbf {\bibinfo {volume} {36}},\ \bibinfo {pages} {125} (\bibinfo {year} {1991})}\BibitemShut {NoStop}%
\bibitem [{Sup()}]{SuppMater}%
  \BibitemOpen
  \href@noop {} {}\bibinfo {note} {See Supplemental Material for additional details regarding the sXRD refinement methodology and results, EDS micrographs showing sample homogeneity and Y-site element mixing, and a comparison of transport behavior and variability within a sample series.}\BibitemShut {Stop}%
\bibitem [{\citenamefont {Gurvitch}\ and\ \citenamefont {Fiory}(1987)}]{gurvitch_resistivity_1987}%
  \BibitemOpen
  \bibfield  {author} {\bibinfo {author} {\bibfnamefont {M.}~\bibnamefont {Gurvitch}}\ and\ \bibinfo {author} {\bibfnamefont {A.~T.}\ \bibnamefont {Fiory}},\ }\href {\doibase 10.1103/PhysRevLett.59.1337} {\bibfield  {journal} {\bibinfo  {journal} {Phys. Rev. Lett.}\ }\textbf {\bibinfo {volume} {59}},\ \bibinfo {pages} {1337} (\bibinfo {year} {1987})}\BibitemShut {NoStop}%
\bibitem [{\citenamefont {Keimer}\ \emph {et~al.}(2015)\citenamefont {Keimer}, \citenamefont {Kivelson}, \citenamefont {Norman}, \citenamefont {Uchida},\ and\ \citenamefont {Zaanen}}]{keimer_quantum_2015}%
  \BibitemOpen
  \bibfield  {author} {\bibinfo {author} {\bibfnamefont {B.}~\bibnamefont {Keimer}}, \bibinfo {author} {\bibfnamefont {S.~A.}\ \bibnamefont {Kivelson}}, \bibinfo {author} {\bibfnamefont {M.~R.}\ \bibnamefont {Norman}}, \bibinfo {author} {\bibfnamefont {S.}~\bibnamefont {Uchida}}, \ and\ \bibinfo {author} {\bibfnamefont {J.}~\bibnamefont {Zaanen}},\ }\href {\doibase 10.1038/nature14165} {\bibfield  {journal} {\bibinfo  {journal} {Nature}\ }\textbf {\bibinfo {volume} {518}},\ \bibinfo {pages} {179} (\bibinfo {year} {2015})}\BibitemShut {NoStop}%
\bibitem [{\citenamefont {Le~Tacon}(2021)}]{le_tacon_strange_2021}%
  \BibitemOpen
  \bibfield  {author} {\bibinfo {author} {\bibfnamefont {M.}~\bibnamefont {Le~Tacon}},\ }\href {\doibase 10.1126/science.abi9685} {\bibfield  {journal} {\bibinfo  {journal} {Science}\ }\textbf {\bibinfo {volume} {373}},\ \bibinfo {pages} {1438} (\bibinfo {year} {2021})}\BibitemShut {NoStop}%
\bibitem [{\citenamefont {Phillips}\ \emph {et~al.}(2022)\citenamefont {Phillips}, \citenamefont {Hussey},\ and\ \citenamefont {Abbamonte}}]{phillips_stranger_2022}%
  \BibitemOpen
  \bibfield  {author} {\bibinfo {author} {\bibfnamefont {P.~W.}\ \bibnamefont {Phillips}}, \bibinfo {author} {\bibfnamefont {N.~E.}\ \bibnamefont {Hussey}}, \ and\ \bibinfo {author} {\bibfnamefont {P.}~\bibnamefont {Abbamonte}},\ }\href {\doibase 10.1126/science.abh4273} {\bibfield  {journal} {\bibinfo  {journal} {Science}\ }\textbf {\bibinfo {volume} {377}},\ \bibinfo {pages} {eabh4273} (\bibinfo {year} {2022})}\BibitemShut {NoStop}%
\bibitem [{\citenamefont {Langan}\ \emph {et~al.}(1998)\citenamefont {Langan}, \citenamefont {Gordeev}, \citenamefont {Jansen}, \citenamefont {Gagnon},\ and\ \citenamefont {Taillefer}}]{langan_phase_1998}%
  \BibitemOpen
  \bibfield  {author} {\bibinfo {author} {\bibfnamefont {R.~M.}\ \bibnamefont {Langan}}, \bibinfo {author} {\bibfnamefont {S.~N.}\ \bibnamefont {Gordeev}}, \bibinfo {author} {\bibfnamefont {A.~G.~M.}\ \bibnamefont {Jansen}}, \bibinfo {author} {\bibfnamefont {R.}~\bibnamefont {Gagnon}}, \ and\ \bibinfo {author} {\bibfnamefont {L.}~\bibnamefont {Taillefer}},\ }\href@noop {} {\bibfield  {journal} {\bibinfo  {journal} {Phys. Rev. B}\ }\textbf {\bibinfo {volume} {58}} (\bibinfo {year} {1998})}\BibitemShut {NoStop}%
\bibitem [{\citenamefont {Graser}\ \emph {et~al.}(2007)\citenamefont {Graser}, \citenamefont {Hirschfeld}, \citenamefont {Zhu},\ and\ \citenamefont {Dahm}}]{graser_t_2007}%
  \BibitemOpen
  \bibfield  {author} {\bibinfo {author} {\bibfnamefont {S.}~\bibnamefont {Graser}}, \bibinfo {author} {\bibfnamefont {P.~J.}\ \bibnamefont {Hirschfeld}}, \bibinfo {author} {\bibfnamefont {L.-Y.}\ \bibnamefont {Zhu}}, \ and\ \bibinfo {author} {\bibfnamefont {T.}~\bibnamefont {Dahm}},\ }\href {\doibase 10.1103/PhysRevB.76.054516} {\bibfield  {journal} {\bibinfo  {journal} {Phys. Rev. B}\ }\textbf {\bibinfo {volume} {76}},\ \bibinfo {pages} {054516} (\bibinfo {year} {2007})}\BibitemShut {NoStop}%
\bibitem [{\citenamefont {Özdemir}\ \emph {et~al.}(2022)\citenamefont {Özdemir}, \citenamefont {Mishra}, \citenamefont {Lee-Hone}, \citenamefont {Kong}, \citenamefont {Berlijn}, \citenamefont {Broun},\ and\ \citenamefont {Hirschfeld}}]{ozdemir_effect_2022}%
  \BibitemOpen
  \bibfield  {author} {\bibinfo {author} {\bibfnamefont {H.~U.}\ \bibnamefont {Özdemir}}, \bibinfo {author} {\bibfnamefont {V.}~\bibnamefont {Mishra}}, \bibinfo {author} {\bibfnamefont {N.~R.}\ \bibnamefont {Lee-Hone}}, \bibinfo {author} {\bibfnamefont {X.}~\bibnamefont {Kong}}, \bibinfo {author} {\bibfnamefont {T.}~\bibnamefont {Berlijn}}, \bibinfo {author} {\bibfnamefont {D.~M.}\ \bibnamefont {Broun}}, \ and\ \bibinfo {author} {\bibfnamefont {P.~J.}\ \bibnamefont {Hirschfeld}},\ }\href {\doibase 10.1103/PhysRevB.106.184510} {\bibfield  {journal} {\bibinfo  {journal} {Phys. Rev. B}\ }\textbf {\bibinfo {volume} {106}},\ \bibinfo {pages} {184510} (\bibinfo {year} {2022})}\BibitemShut {NoStop}%
\bibitem [{\citenamefont {Devi}\ \emph {et~al.}(2000)\citenamefont {Devi}, \citenamefont {Bai}, \citenamefont {Patanjali}, \citenamefont {Pinto}, \citenamefont {Kumar},\ and\ \citenamefont {Malik}}]{devi_enhanced_2000}%
  \BibitemOpen
  \bibfield  {author} {\bibinfo {author} {\bibfnamefont {A.~R.}\ \bibnamefont {Devi}}, \bibinfo {author} {\bibfnamefont {V.~S.}\ \bibnamefont {Bai}}, \bibinfo {author} {\bibfnamefont {P.~V.}\ \bibnamefont {Patanjali}}, \bibinfo {author} {\bibfnamefont {R.}~\bibnamefont {Pinto}}, \bibinfo {author} {\bibfnamefont {N.~H.}\ \bibnamefont {Kumar}}, \ and\ \bibinfo {author} {\bibfnamefont {S.~K.}\ \bibnamefont {Malik}},\ }\href {\doibase 10.1088/0953-2048/13/7/305} {\bibfield  {journal} {\bibinfo  {journal} {Supercond. Sci. Technol.}\ }\textbf {\bibinfo {volume} {13}},\ \bibinfo {pages} {935} (\bibinfo {year} {2000})}\BibitemShut {NoStop}%
\bibitem [{\citenamefont {MacManus-Driscoll}\ \emph {et~al.}(2005)\citenamefont {MacManus-Driscoll}, \citenamefont {Foltyn}, \citenamefont {Maiorov}, \citenamefont {Jia}, \citenamefont {Wang}, \citenamefont {Serquis}, \citenamefont {Civale}, \citenamefont {Lin}, \citenamefont {Hawley}, \citenamefont {Maley},\ and\ \citenamefont {Peterson}}]{macmanus-driscoll_rare_2005}%
  \BibitemOpen
  \bibfield  {author} {\bibinfo {author} {\bibfnamefont {J.~L.}\ \bibnamefont {MacManus-Driscoll}}, \bibinfo {author} {\bibfnamefont {S.~R.}\ \bibnamefont {Foltyn}}, \bibinfo {author} {\bibfnamefont {B.}~\bibnamefont {Maiorov}}, \bibinfo {author} {\bibfnamefont {Q.~X.}\ \bibnamefont {Jia}}, \bibinfo {author} {\bibfnamefont {H.}~\bibnamefont {Wang}}, \bibinfo {author} {\bibfnamefont {A.}~\bibnamefont {Serquis}}, \bibinfo {author} {\bibfnamefont {L.}~\bibnamefont {Civale}}, \bibinfo {author} {\bibfnamefont {Y.}~\bibnamefont {Lin}}, \bibinfo {author} {\bibfnamefont {M.~E.}\ \bibnamefont {Hawley}}, \bibinfo {author} {\bibfnamefont {M.~P.}\ \bibnamefont {Maley}}, \ and\ \bibinfo {author} {\bibfnamefont {D.~E.}\ \bibnamefont {Peterson}},\ }\href {\doibase 10.1063/1.1851006} {\bibfield  {journal} {\bibinfo  {journal} {Appl. Phys. Lett.}\ }\textbf {\bibinfo {volume} {86}},\ \bibinfo {pages} {032505} (\bibinfo {year} {2005})}\BibitemShut {NoStop}%
\bibitem [{\citenamefont {Haugan}\ \emph {et~al.}(2008)\citenamefont {Haugan}, \citenamefont {Campbell}, \citenamefont {Pierce}, \citenamefont {Locke}, \citenamefont {Maartense},\ and\ \citenamefont {Barnes}}]{haugan_microstructural_2008}%
  \BibitemOpen
  \bibfield  {author} {\bibinfo {author} {\bibfnamefont {T.~J.}\ \bibnamefont {Haugan}}, \bibinfo {author} {\bibfnamefont {T.~A.}\ \bibnamefont {Campbell}}, \bibinfo {author} {\bibfnamefont {N.~A.}\ \bibnamefont {Pierce}}, \bibinfo {author} {\bibfnamefont {M.~F.}\ \bibnamefont {Locke}}, \bibinfo {author} {\bibfnamefont {I.}~\bibnamefont {Maartense}}, \ and\ \bibinfo {author} {\bibfnamefont {P.~N.}\ \bibnamefont {Barnes}},\ }\href {\doibase 10.1088/0953-2048/21/2/025014} {\bibfield  {journal} {\bibinfo  {journal} {Supercond. Sci. Technol.}\ }\textbf {\bibinfo {volume} {21}},\ \bibinfo {pages} {025014} (\bibinfo {year} {2008})}\BibitemShut {NoStop}%
\bibitem [{\citenamefont {Wee}\ \emph {et~al.}(2011)\citenamefont {Wee}, \citenamefont {Specht}, \citenamefont {Cantoni}, \citenamefont {Zuev}, \citenamefont {Maroni}, \citenamefont {Wong-Ng}, \citenamefont {Liu}, \citenamefont {Haugan},\ and\ \citenamefont {Goyal}}]{wee_formation_2011}%
  \BibitemOpen
  \bibfield  {author} {\bibinfo {author} {\bibfnamefont {S.~H.}\ \bibnamefont {Wee}}, \bibinfo {author} {\bibfnamefont {E.~D.}\ \bibnamefont {Specht}}, \bibinfo {author} {\bibfnamefont {C.}~\bibnamefont {Cantoni}}, \bibinfo {author} {\bibfnamefont {Y.~L.}\ \bibnamefont {Zuev}}, \bibinfo {author} {\bibfnamefont {V.}~\bibnamefont {Maroni}}, \bibinfo {author} {\bibfnamefont {W.}~\bibnamefont {Wong-Ng}}, \bibinfo {author} {\bibfnamefont {G.}~\bibnamefont {Liu}}, \bibinfo {author} {\bibfnamefont {T.~J.}\ \bibnamefont {Haugan}}, \ and\ \bibinfo {author} {\bibfnamefont {A.}~\bibnamefont {Goyal}},\ }\href {\doibase 10.1103/PhysRevB.83.224520} {\bibfield  {journal} {\bibinfo  {journal} {Phys. Rev. B}\ }\textbf {\bibinfo {volume} {83}},\ \bibinfo {pages} {224520} (\bibinfo {year} {2011})}\BibitemShut {NoStop}%
\bibitem [{\citenamefont {Tuilier}\ \emph {et~al.}(1992)\citenamefont {Tuilier}, \citenamefont {Chevalier}, \citenamefont {Tressaud}, \citenamefont {Brisson}, \citenamefont {Soubeyroux},\ and\ \citenamefont {Etourneau}}]{tuilier1992exafs}%
  \BibitemOpen
  \bibfield  {author} {\bibinfo {author} {\bibfnamefont {M.}~\bibnamefont {Tuilier}}, \bibinfo {author} {\bibfnamefont {B.}~\bibnamefont {Chevalier}}, \bibinfo {author} {\bibfnamefont {A.}~\bibnamefont {Tressaud}}, \bibinfo {author} {\bibfnamefont {C.}~\bibnamefont {Brisson}}, \bibinfo {author} {\bibfnamefont {J.}~\bibnamefont {Soubeyroux}}, \ and\ \bibinfo {author} {\bibfnamefont {J.}~\bibnamefont {Etourneau}},\ }\href@noop {} {\bibfield  {journal} {\bibinfo  {journal} {Physica C}\ }\textbf {\bibinfo {volume} {200}},\ \bibinfo {pages} {113} (\bibinfo {year} {1992})}\BibitemShut {NoStop}%
\bibitem [{\citenamefont {Wakimoto}\ \emph {et~al.}(2004)\citenamefont {Wakimoto}, \citenamefont {Zhang}, \citenamefont {Yamada}, \citenamefont {Swainson}, \citenamefont {Kim},\ and\ \citenamefont {Birgeneau}}]{wakimoto_direct_2004}%
  \BibitemOpen
  \bibfield  {author} {\bibinfo {author} {\bibfnamefont {S.}~\bibnamefont {Wakimoto}}, \bibinfo {author} {\bibfnamefont {H.}~\bibnamefont {Zhang}}, \bibinfo {author} {\bibfnamefont {K.}~\bibnamefont {Yamada}}, \bibinfo {author} {\bibfnamefont {I.}~\bibnamefont {Swainson}}, \bibinfo {author} {\bibfnamefont {H.}~\bibnamefont {Kim}}, \ and\ \bibinfo {author} {\bibfnamefont {R.~J.}\ \bibnamefont {Birgeneau}},\ }\href {\doibase 10.1103/PhysRevLett.92.217004} {\bibfield  {journal} {\bibinfo  {journal} {Phys. Rev. Lett.}\ }\textbf {\bibinfo {volume} {92}},\ \bibinfo {pages} {217004} (\bibinfo {year} {2004})}\BibitemShut {NoStop}%
\bibitem [{\citenamefont {Le~Tacon}\ \emph {et~al.}(2011)\citenamefont {Le~Tacon}, \citenamefont {Ghiringhelli}, \citenamefont {Chaloupka}, \citenamefont {Sala}, \citenamefont {Hinkov}, \citenamefont {Haverkort}, \citenamefont {Minola}, \citenamefont {Bakr}, \citenamefont {Zhou}, \citenamefont {Blanco-Canosa}, \citenamefont {Monney}, \citenamefont {Song}, \citenamefont {Sun}, \citenamefont {Lin}, \citenamefont {De~Luca}, \citenamefont {Salluzzo}, \citenamefont {Khaliullin}, \citenamefont {Schmitt}, \citenamefont {Braicovich},\ and\ \citenamefont {Keimer}}]{le_tacon_intense_2011}%
  \BibitemOpen
  \bibfield  {author} {\bibinfo {author} {\bibfnamefont {M.}~\bibnamefont {Le~Tacon}}, \bibinfo {author} {\bibfnamefont {G.}~\bibnamefont {Ghiringhelli}}, \bibinfo {author} {\bibfnamefont {J.}~\bibnamefont {Chaloupka}}, \bibinfo {author} {\bibfnamefont {M.~M.}\ \bibnamefont {Sala}}, \bibinfo {author} {\bibfnamefont {V.}~\bibnamefont {Hinkov}}, \bibinfo {author} {\bibfnamefont {M.~W.}\ \bibnamefont {Haverkort}}, \bibinfo {author} {\bibfnamefont {M.}~\bibnamefont {Minola}}, \bibinfo {author} {\bibfnamefont {M.}~\bibnamefont {Bakr}}, \bibinfo {author} {\bibfnamefont {K.~J.}\ \bibnamefont {Zhou}}, \bibinfo {author} {\bibfnamefont {S.}~\bibnamefont {Blanco-Canosa}}, \bibinfo {author} {\bibfnamefont {C.}~\bibnamefont {Monney}}, \bibinfo {author} {\bibfnamefont {Y.~T.}\ \bibnamefont {Song}}, \bibinfo {author} {\bibfnamefont {G.~L.}\ \bibnamefont {Sun}}, \bibinfo {author} {\bibfnamefont {C.~T.}\ \bibnamefont {Lin}}, \bibinfo {author} {\bibfnamefont {G.~M.}\ \bibnamefont {De~Luca}}, \bibinfo {author} {\bibfnamefont
  {M.}~\bibnamefont {Salluzzo}}, \bibinfo {author} {\bibfnamefont {G.}~\bibnamefont {Khaliullin}}, \bibinfo {author} {\bibfnamefont {T.}~\bibnamefont {Schmitt}}, \bibinfo {author} {\bibfnamefont {L.}~\bibnamefont {Braicovich}}, \ and\ \bibinfo {author} {\bibfnamefont {B.}~\bibnamefont {Keimer}},\ }\href {\doibase 10.1038/nphys2041} {\bibfield  {journal} {\bibinfo  {journal} {Nature Phys}\ }\textbf {\bibinfo {volume} {7}},\ \bibinfo {pages} {725} (\bibinfo {year} {2011})}\BibitemShut {NoStop}%
\bibitem [{\citenamefont {Scalapino}(2012)}]{scalapino_common_2012}%
  \BibitemOpen
  \bibfield  {author} {\bibinfo {author} {\bibfnamefont {D.~J.}\ \bibnamefont {Scalapino}},\ }\href {\doibase 10.1103/RevModPhys.84.1383} {\bibfield  {journal} {\bibinfo  {journal} {Rev. Mod. Phys.}\ }\textbf {\bibinfo {volume} {84}},\ \bibinfo {pages} {1383} (\bibinfo {year} {2012})}\BibitemShut {NoStop}%
\bibitem [{\citenamefont {Proust}\ and\ \citenamefont {Taillefer}(2019)}]{proust_remarkable_2019}%
  \BibitemOpen
  \bibfield  {author} {\bibinfo {author} {\bibfnamefont {C.}~\bibnamefont {Proust}}\ and\ \bibinfo {author} {\bibfnamefont {L.}~\bibnamefont {Taillefer}},\ }\href {\doibase 10.1146/annurev-conmatphys-031218-013210} {\bibfield  {journal} {\bibinfo  {journal} {Annu. Rev. Condens. Matter Phys.}\ }\textbf {\bibinfo {volume} {10}},\ \bibinfo {pages} {409} (\bibinfo {year} {2019})}\BibitemShut {NoStop}%
\bibitem [{\citenamefont {Lanzara}\ \emph {et~al.}(2001)\citenamefont {Lanzara}, \citenamefont {Bogdanov}, \citenamefont {Zhou}, \citenamefont {Kellar}, \citenamefont {Feng}, \citenamefont {Lu}, \citenamefont {Yoshida}, \citenamefont {Eisaki}, \citenamefont {Fujimori}, \citenamefont {Kishio}, \citenamefont {Shimoyama}, \citenamefont {Noda}, \citenamefont {Uchida}, \citenamefont {Hussain},\ and\ \citenamefont {Shen}}]{lanzara_evidence_2001}%
  \BibitemOpen
  \bibfield  {author} {\bibinfo {author} {\bibfnamefont {A.}~\bibnamefont {Lanzara}}, \bibinfo {author} {\bibfnamefont {P.~V.}\ \bibnamefont {Bogdanov}}, \bibinfo {author} {\bibfnamefont {X.~J.}\ \bibnamefont {Zhou}}, \bibinfo {author} {\bibfnamefont {S.~A.}\ \bibnamefont {Kellar}}, \bibinfo {author} {\bibfnamefont {D.~L.}\ \bibnamefont {Feng}}, \bibinfo {author} {\bibfnamefont {E.~D.}\ \bibnamefont {Lu}}, \bibinfo {author} {\bibfnamefont {T.}~\bibnamefont {Yoshida}}, \bibinfo {author} {\bibfnamefont {H.}~\bibnamefont {Eisaki}}, \bibinfo {author} {\bibfnamefont {A.}~\bibnamefont {Fujimori}}, \bibinfo {author} {\bibfnamefont {K.}~\bibnamefont {Kishio}}, \bibinfo {author} {\bibfnamefont {J.-I.}\ \bibnamefont {Shimoyama}}, \bibinfo {author} {\bibfnamefont {T.}~\bibnamefont {Noda}}, \bibinfo {author} {\bibfnamefont {S.}~\bibnamefont {Uchida}}, \bibinfo {author} {\bibfnamefont {Z.}~\bibnamefont {Hussain}}, \ and\ \bibinfo {author} {\bibfnamefont {Z.-X.}\ \bibnamefont {Shen}},\ }\href {\doibase 10.1038/35087518}
  {\bibfield  {journal} {\bibinfo  {journal} {Nature}\ }\textbf {\bibinfo {volume} {412}},\ \bibinfo {pages} {510} (\bibinfo {year} {2001})}\BibitemShut {NoStop}%
\bibitem [{\citenamefont {Iwasawa}\ \emph {et~al.}(2008)\citenamefont {Iwasawa}, \citenamefont {Douglas}, \citenamefont {Sato}, \citenamefont {Masui}, \citenamefont {Yoshida}, \citenamefont {Sun}, \citenamefont {Eisaki}, \citenamefont {Bando}, \citenamefont {Ino}, \citenamefont {Arita}, \citenamefont {Shimada}, \citenamefont {Namatame}, \citenamefont {Taniguchi}, \citenamefont {Tajima}, \citenamefont {Uchida}, \citenamefont {Saitoh}, \citenamefont {Dessau},\ and\ \citenamefont {Aiura}}]{iwasawa_isotopic_2008}%
  \BibitemOpen
  \bibfield  {author} {\bibinfo {author} {\bibfnamefont {H.}~\bibnamefont {Iwasawa}}, \bibinfo {author} {\bibfnamefont {J.~F.}\ \bibnamefont {Douglas}}, \bibinfo {author} {\bibfnamefont {K.}~\bibnamefont {Sato}}, \bibinfo {author} {\bibfnamefont {T.}~\bibnamefont {Masui}}, \bibinfo {author} {\bibfnamefont {Y.}~\bibnamefont {Yoshida}}, \bibinfo {author} {\bibfnamefont {Z.}~\bibnamefont {Sun}}, \bibinfo {author} {\bibfnamefont {H.}~\bibnamefont {Eisaki}}, \bibinfo {author} {\bibfnamefont {H.}~\bibnamefont {Bando}}, \bibinfo {author} {\bibfnamefont {A.}~\bibnamefont {Ino}}, \bibinfo {author} {\bibfnamefont {M.}~\bibnamefont {Arita}}, \bibinfo {author} {\bibfnamefont {K.}~\bibnamefont {Shimada}}, \bibinfo {author} {\bibfnamefont {H.}~\bibnamefont {Namatame}}, \bibinfo {author} {\bibfnamefont {M.}~\bibnamefont {Taniguchi}}, \bibinfo {author} {\bibfnamefont {S.}~\bibnamefont {Tajima}}, \bibinfo {author} {\bibfnamefont {S.}~\bibnamefont {Uchida}}, \bibinfo {author} {\bibfnamefont {T.}~\bibnamefont {Saitoh}},
  \bibinfo {author} {\bibfnamefont {D.~S.}\ \bibnamefont {Dessau}}, \ and\ \bibinfo {author} {\bibfnamefont {Y.}~\bibnamefont {Aiura}},\ }\href {\doibase 10.1103/PhysRevLett.101.157005} {\bibfield  {journal} {\bibinfo  {journal} {Phys. Rev. Lett.}\ }\textbf {\bibinfo {volume} {101}},\ \bibinfo {pages} {157005} (\bibinfo {year} {2008})}\BibitemShut {NoStop}%
\bibitem [{\citenamefont {Lee}\ \emph {et~al.}(2006)\citenamefont {Lee}, \citenamefont {Fujita}, \citenamefont {McElroy}, \citenamefont {Slezak}, \citenamefont {Wang}, \citenamefont {Aiura}, \citenamefont {Bando}, \citenamefont {Ishikado}, \citenamefont {Masui}, \citenamefont {Zhu}, \citenamefont {Balatsky}, \citenamefont {Eisaki}, \citenamefont {Uchida},\ and\ \citenamefont {Davis}}]{lee_interplay_2006}%
  \BibitemOpen
  \bibfield  {author} {\bibinfo {author} {\bibfnamefont {J.}~\bibnamefont {Lee}}, \bibinfo {author} {\bibfnamefont {K.}~\bibnamefont {Fujita}}, \bibinfo {author} {\bibfnamefont {K.}~\bibnamefont {McElroy}}, \bibinfo {author} {\bibfnamefont {J.~A.}\ \bibnamefont {Slezak}}, \bibinfo {author} {\bibfnamefont {M.}~\bibnamefont {Wang}}, \bibinfo {author} {\bibfnamefont {Y.}~\bibnamefont {Aiura}}, \bibinfo {author} {\bibfnamefont {H.}~\bibnamefont {Bando}}, \bibinfo {author} {\bibfnamefont {M.}~\bibnamefont {Ishikado}}, \bibinfo {author} {\bibfnamefont {T.}~\bibnamefont {Masui}}, \bibinfo {author} {\bibfnamefont {J.-X.}\ \bibnamefont {Zhu}}, \bibinfo {author} {\bibfnamefont {A.~V.}\ \bibnamefont {Balatsky}}, \bibinfo {author} {\bibfnamefont {H.}~\bibnamefont {Eisaki}}, \bibinfo {author} {\bibfnamefont {S.}~\bibnamefont {Uchida}}, \ and\ \bibinfo {author} {\bibfnamefont {J.~C.}\ \bibnamefont {Davis}},\ }\href {\doibase 10.1038/nature04973} {\bibfield  {journal} {\bibinfo  {journal} {Nature}\ }\textbf {\bibinfo
  {volume} {442}},\ \bibinfo {pages} {546} (\bibinfo {year} {2006})}\BibitemShut {NoStop}%
\end{thebibliography}%

\end{document}